\documentclass[aps,twocolumn,amsfonts,amssymb,longbibliography,superscriptaddress]{revtex4-1}
\usepackage{graphicx}
\usepackage{multirow}
\usepackage{epstopdf} 
\usepackage{dcolumn}
\usepackage{bm}
\usepackage{bbm,bbold}
\usepackage{color}
\usepackage{xcolor, soul}
\sethlcolor{green}
\usepackage{amsfonts}
\usepackage{amsmath}
\usepackage{mdframed}
\usepackage[normalem]{ulem}
\usepackage{mathrsfs}   
\usepackage[none]{hyphenat}
\usepackage{subfigure}
\usepackage{float}
\usepackage{physics}
\usepackage[colorlinks=true,citecolor=blue]{hyperref}
\hypersetup{colorlinks=true,citecolor=blue,linkcolor=blue,urlcolor=blue}

\begin{document}
	\title{Disorder induced dynamical interband response in Dirac nodal line semimetals}	
	\author{Vivek Pandey}
    \email{vivek_pandey@srmap.edu.in}
	\affiliation{Department of Physics, School of Engineering and Sciences, SRM University AP, Amaravati, 522240, India}
 	\author{Pankaj Bhalla}
	\email{pankaj.b@srmap.edu.in}
 \affiliation{Department of Physics, School of Engineering and Sciences, SRM University AP, Amaravati, 522240, India}
    
\date{\today}

\begin{abstract}
To obtain the total response of the system, the effect of disorder cannot be neglected, as it introduces a new contribution (i.e. extrinsic) in the total response of the system. In the study of dynamical (AC) effects, the interband response exhibits an exotic resonance peak due to interband transitions. Here, the dynamical interband response of Dirac nodal line semimetal is investigated by using the quantum kinetic approach. 
The scattering driven effect is analyzed under the first-order Born approximation (i.e., in the weak disorder limit) and reveals a resonance peak at $2\tilde{\mu}$. In contrast, the field driven intrinsic response peak depends on both the mass ($\tilde{M}$) and chemical potential ($\tilde{\mu}$). 
The results indicate that the total interband response of the 3D nodal line semimetals, is mainly dominated by the disorder induced contributions.

\end{abstract}

\maketitle

\section{Introduction}

The topological semimetals can be categorized with the presence of a nodal point or a nodal line. The nodal line semimetals (NLSMs) are a type of topological semimetal which form due to the band overlapping between the conduction and valence bands either as a one-dimensional line or a loop in the Brillouin zone~\cite{Burkov_prb2011, Bian_nature2016, Shuo_apx2018, Bian_prb2016, Fang_prb2015, Yu_prl2015, Xie_APLM2015, Kim_prl2015, Ekahana_njp2017, Feng_prm2018, schoop_nc2016, Neupane_prb2016, Hu_prl2016, Takane_prb2017, Chen_prb2017, Wang_prb2017, Liang_prb2016, Fang_cpb2016, Rui_prb2018, yu_fop2017, Xu_prb2017}.
The nodal line semimetal is more robust towards the perturbation due to the presence of the continuous band touching compared to Dirac~\cite{Wang_prb2013, liu_nm2014, Borisenko_prl2014, yi_sr2014, liu_science2014, xu_science2015}  and Weyl~\cite{Lv_prx2015, Weng_prx2015, xu_nc2016, Belopolski_prl2016, Souma_prb2016, Xu_sa2015} semimetals where band touching happens in a point (i.e. nodal point) or a pair of nodal points.
Further the nodal line semimetal includes Dirac nodal line semimetals (DNLSMs)~\cite{Rui_prb2018}, which are symmetry protected and fourfold degenerate and Weyl nodal line semimetals (WNLSMs), having either inversion ($\mathcal{P}$) symmetry or time reversal ($\mathcal{T}$) symmetry broken. Some materials under DNLSMs are Ca$_3$P$_2$~\cite{Xie_APLM2015, Chan_prb2016}, Cu$_3$N~\cite{Kim_prl2015}, PtSn$_4$~\cite{Lin_nl2024},  ZrSiS~\cite{Fu_sa2019}, CaAgP and CaAgAs~\cite{Okamoto_jpsj2016} and under WNLSMs are LaNiSi, LaPtSi and LaPtGe materials~\cite{shang_npj2022}. Notably,  Ca$_3$P$_2$ has a band touching (nodal ring) occurring exactly at the Fermi energy. However, other materials show similar features on tuning the Fermi energy via doping~\cite{Chan_prb2016, Xie_APLM2015}. Hence, Ca$_3$P$_2$ is an ideal testing material to make thorough investigation about the transport properties of nodal line semimetals within the developed theoretical framework~\cite{Barati_prb2017}.

\begin{figure}[t]
    \centering
    \includegraphics[width=\linewidth]{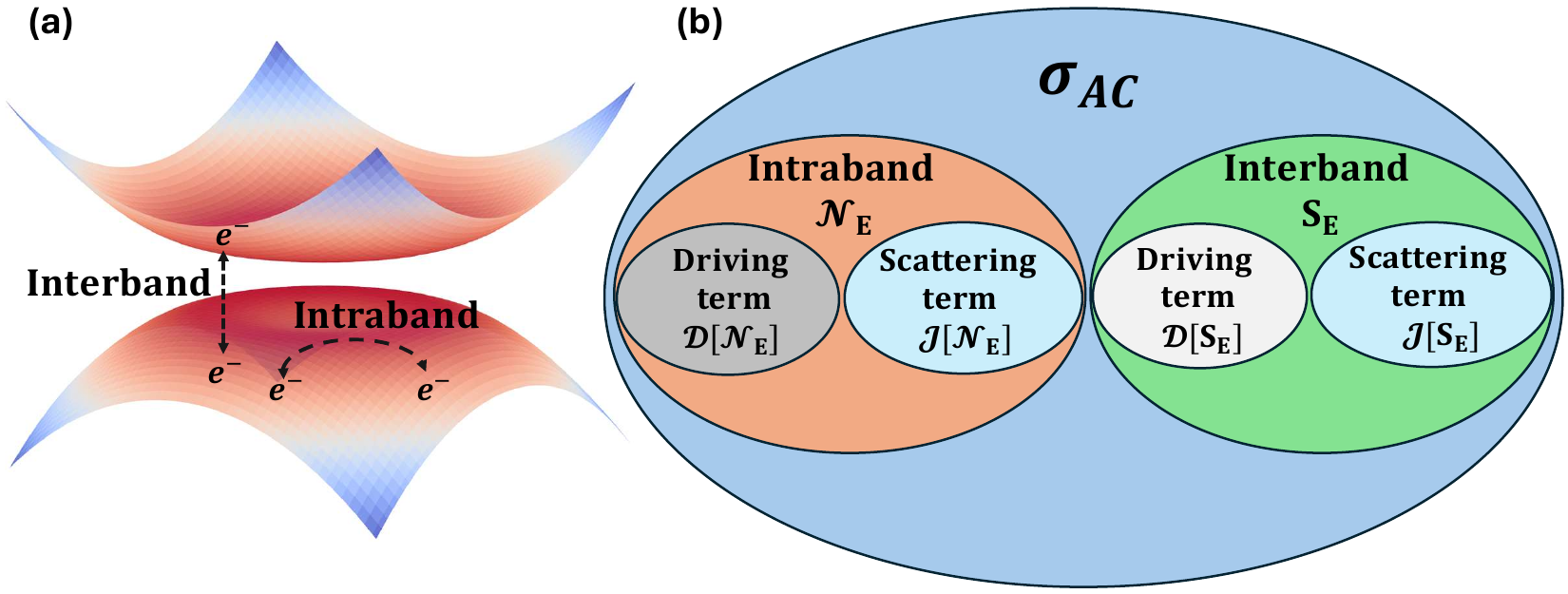}
    \caption{(a) represents the schematic diagram which shows the intraband and interband transition, and (b) depicts a venn diagram of the AC conductivity ($\sigma_{AC}$), where $\sigma_{AC}$ comes from the field correction terms to the distribution function namely the intraband ($\mathcal{N}_{E}$) and the interband ($S_E$). Additionally, the contributions to intraband and interband responses come from the field driving term ($\mathcal{D}[\mathcal{N}_E]$ or $\mathcal{D}[S_E]$) and scattering driving term ($\mathcal{J}[\mathcal{N}_E]$ or $\mathcal{J}[S_E]$) respectively.} 
    \label{fig:Schematic}
\end{figure}

To deal with the transport, the quantum kinetic approach is employed which provides the immediate separation of intrinsic effects, extrinsic effects, and effects that combine interband coherence and disorder.~\cite{Culcer_prb2017}.  
These effects lead to the interband and intraband contributions to the conductivity of a system. In figure~\ref{fig:Schematic}(a), the schematic view of the intraband that involves only single-band dynamics and interband which contains the transitions between two distinct bands is provided.
Further, the interband and intraband originate via two contributions, the first arises from the Bloch band and shows the field dependent response known as the intrinsic response. The second part arises from the effect of the disorder on the system response and is known as the extrinsic response as shown in figure~\ref{fig:Schematic}(b).
Unlike intraband response, which gives a significant result mainly within the DC limit, the interband response dictates interesting features over a broad frequency range such as well-defined resonances or absorption peaks. Interestingly, these features can be tuned by chemical potential shifts, band gap engineering and external perturbations which makes the interband contribution important for material prospects.

\begin{figure*}[htp]
    \centering
    \includegraphics[width=15cm]{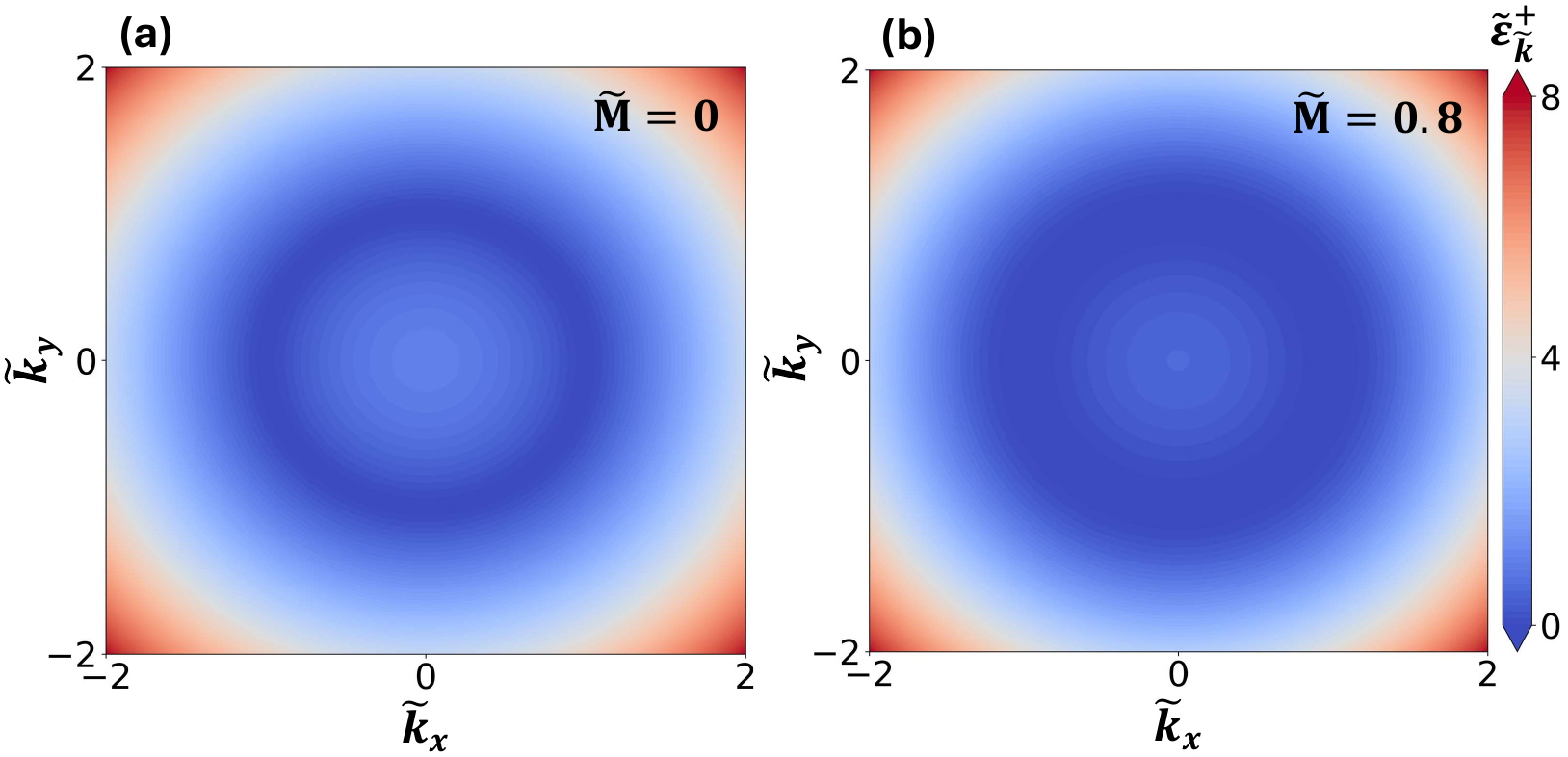}
    \caption{(a) and (b) shows contour plot for the energy dispersion ($\tilde{\varepsilon}_{\bm k}^{+}$) at normalized mass values $\tilde{M}=0$ and $\tilde{M}=0.8$ with respect to $\varepsilon_0$, an energy associated with nodal ring radius. Here, $\tilde{k}_x$ and $\tilde{k}_y$ refer to the normalized wave vectors with respect to the nodal ring radius $k_0$ in the $x$ and $y$ directions respectively.} 
    \label{fig:1}
\end{figure*}
Recently, the interband conductivity for the case of Weyl and inverted band semimetals are investigated by Rukelj et al.~\cite{Rukelj_prb2020, Rukelj_prb2021,Rukelj_prb2020_m}. The authors found that the optical interband conductivity for Weyl semimetals vanishes at Weyl nodes due to vanishing density of states and shows different frequency power laws in high and low frequency region~\cite{Rukelj_prb2020}. In the case of band inverted system, the interband conductivity shows the diverging response (resonance peaks) when the frequency approaches to the band gap due to the diverging DOS~\cite{Rukelj_prb2021}. 
Further, Barati et al. and others have discussed the intrinsic contribution to the interband part for 3D nodal line semimetals~\cite{Barati_prb2017, Wang_prb2021}. In addition, the extrinsic contribution to the dc response is calculated by Pandey et al.~\cite{Pandey_prb2024}, and analyzes the conductivity within dc regimes. However, the understanding of the total interband part (intrinsic and extrinsic) of the response on the application of an oscillating electric field (AC field) is lacking and has not been addressed so far. This motivates the investigation of dynamical response for DNLSMs and finds how the scattering driven effects can amplify the overall dynamical interband response of the system in different frequency regimes. 

The present work focuses on the dynamical longitudinal interband response of the 3D DNLSMs by applying quantum kinetic theory. The central point of this work is the disorder contribution to the dynamical conductivity. We find that the scattering driven interband response (extrinsic) which arises from the effect of random disorder on the system dominates the overall interband response of the system in the lower and mid frequency regime. Additionally, we discuss the tunability of the interband response with the chemical potential ($\mu$), and the mass term ($M$).

\section{Hamiltonian and Theoretical methodology}\label{sec:model}
The gaped DNLSMs around the $\Gamma$ point is described by a low energy effective $\bm {k.p}$ two-band model Hamiltonian having broken $\mathcal{PT}$ symmetry, where $\mathcal{P}$ and $\mathcal{T}$ stand for inversion and time reversal respectively, as  
\begin{equation}\label{eqn:hamitonian}
    \mathcal{H}({\bm k}) =  \frac{\hbar^{2}k_0^2} {2m}\bigg[\space\bigg(\frac{\mathcal{K}}{k_0}-1\bigg) \space\sigma_x +  \space \frac{2m v_z }{\hbar k_0^2} k_z\space\sigma_y + \frac{2m}{\hbar^2 k_0^2}M \space \sigma_z\bigg]. 
\end{equation}
Here $\mathcal{K}=\sqrt{k_x^2+k_y^2}$ is the magnitude of the planar wave vector in $k_x-k_y$ plane, $k_0$ is the nodal ring radius, $v_z$ is the component of Fermi velocity along $z$-direction, $\sigma_{i} (i \equiv x,y,z)$  refers to the Pauli matrices in the pseudo spin basis and $m$ is the electronic mass. $M$ is the mass (gap) term that emerges experimentally from the application of an external field, pressure, stress, etc., and is the source of the broken $\mathcal{PT}$ symmetry of the system~\cite{chiba_prb2017, Kot_prb2020, chen_prb2018, Wang_prb2021, Rendy_jap2021, du_nrp2021, Flores-Calderón_EL2023, Pandey_prb2024}.
Corresponding to the two-band model Hamiltonian equation~\eqref{eqn:hamitonian}, the band dispersion and eigenvectors are
\begin{eqnarray}
   & \tilde{\varepsilon}_{\bm k}^n =  n \sqrt{(\tilde{\mathcal{K}} - 1)^2+(\gamma \tilde{k}_z)^2 + \tilde{M}^2}~;   |u_{{\bm k}}^{n} \rangle = \left(
    \begin{matrix}
        \zeta^{-}_{\bm k} e^{-i\theta_{\tilde{\bm{k}}}} \\
        \zeta^{+}_{\bm k}
    \end{matrix}
    \right).
\end{eqnarray}
Here we consider $ \tilde{\varepsilon}_{\bm k}^n = \varepsilon_{\bm k}^n/\varepsilon_0, \tilde{\mathcal{K}} =\mathcal{K}/k_0, \tilde{k}_z=k_z/k_0, \gamma=2 m v_z/\hbar k_0$, $\tilde{M}=M/\varepsilon_0$, $\varepsilon_0=\hbar^2 k_0^2/(2m)$ and $n $ for two distinct bands such as conduction ($n=+$) band and valence $(n=-)$ band. 
The energy dispersion at zero and the finite mass are shown in figure~\ref{fig:1}.
Further, we take $\zeta_{\bm k}^{\pm}=\pm \frac{1}{\sqrt{2}}[1\pm \tilde{M}/\tilde{\varepsilon}_{\bm k}^{n}]^{1/2}$ and $\theta_{\tilde{\bm{k}}}$ the angle between the $\tilde{\mathcal{K}}$ plane and the $\tilde{k}_z$ direction as $\theta_{\tilde{\bm{k}}} = \tan^{-1} [\gamma \tilde{k}_z/ (\tilde{\mathcal{K}}-1)]$. The low energy band structure is in accord with references~\cite{Xie_APLM2015, Chan_prb2016}, where authors predict the existence of a nodal ring in the $k_x - k_y$ plane for Ca$_3$P$_2$ system.

\begin{figure*}[htp]
    \centering
    \includegraphics[width=15cm]{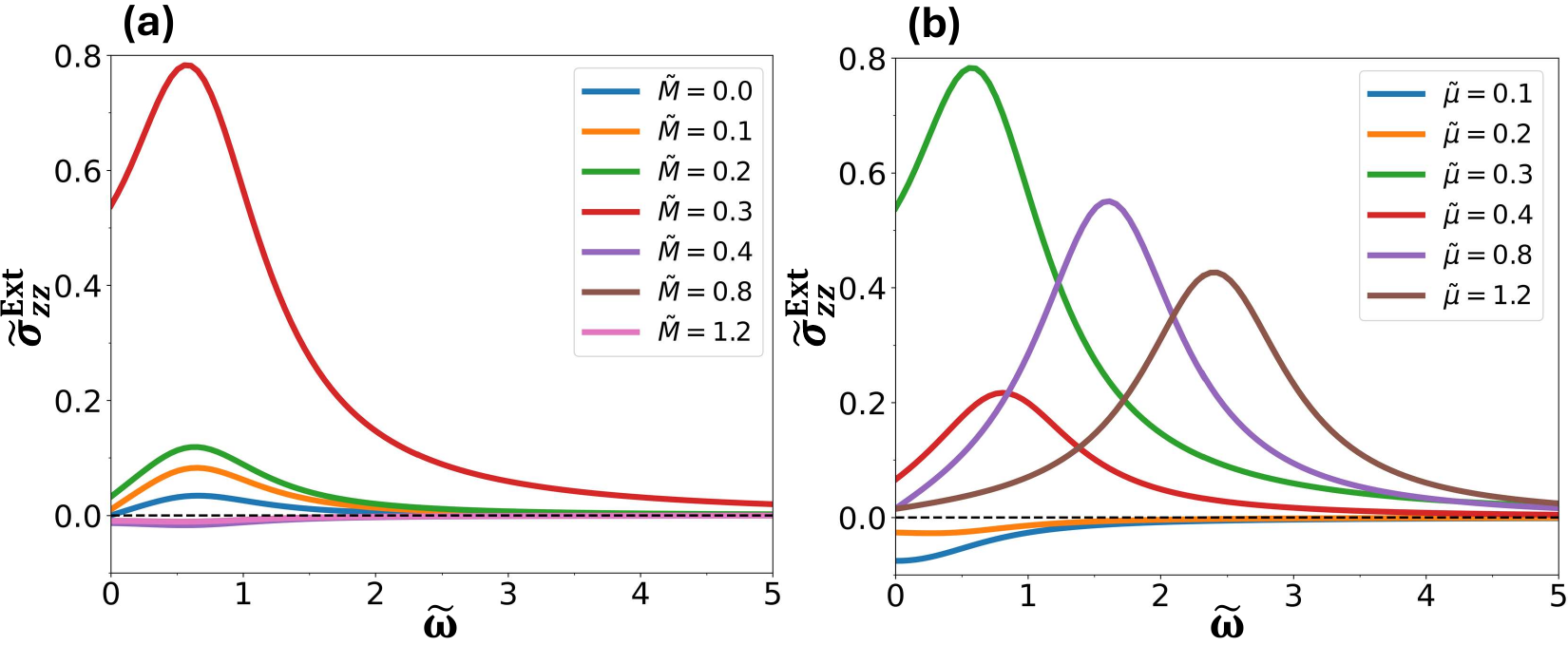}
    \caption{(a) and (b) depict the extrinsic contribution to the interband part of the longitudinal conductivity of DNLSM  ($\tilde{\sigma}_{zz}^{\text{Ext}}=\sigma_{zz}^{\text{Ext}}/\sigma_0$) where $\sigma_0 = e^2 \gamma^2 k_0/\hbar(2 \pi)^2$ as a function of frequency at distinct values of mass (at constant $\tilde{\mu}=0.3$) and chemical potential (at constant $\tilde{M}=0.3$).} 
    \label{fig:Extrinsic}
\end{figure*}

To compute the dynamical response of the system we employ the quantum kinetic approach. The brief details of the latter approach is provided below.

The  quantum Liouville equation for the effective single-particle density matrix $\rho$, is written as~\cite{Culcer_prb2017, Bhalla_prb2023, Pandey_prb2024}
\begin{equation} \label{eqn:4}
    \frac{\partial  \rho}{\partial t}+\frac{i}{\hbar} [ \mathcal{H}_0, \rho]=D_{\bm E, \bm k}[\rho]-\mathcal{J}_{\bm k}[\rho],
\end{equation}
where we consider the full Hamiltonian of the system $\mathcal{H}=\mathcal{H}_0  + \mathcal{H}_{\bm E} + U$, that leads to three terms associated with individual component of $\mathcal{H}$. The first term includes an unperturbed Hamiltonian $\mathcal{H}_0$ that follows the eigenvalue equation $\mathcal{H}_0 \ket{u_{\bm{k}}^n}=\varepsilon_{\bm{k}}^n\ket{u_{\bm{k}}^n}$ where $\ket{u_{\bm{k}}^n}$ is the eigenfunction and $\varepsilon_{\bm{k}}^n$ is the eigenenergy and $n$ is the band index.
%

%
The second term $D_{\bm E, \bm k}[\rho]=-i/\hbar[\mathcal{H}_{\bm E},\rho]$ is embedded with perturbed part of the Hamiltonian $\mathcal{H}_{\bm E} = e \bm{E}(t).\hat{\bm {r}}$ having external electric field $\bm {E}(t)=E e^{i\omega t}$, which is time dependent and spatially homogeneous in nature. The $\hat{\bm r}$ is the position vector attached to the electron with charge $e$. Within the band basis representation $\hbar D_{\bm E,\bm{k}}^{np} =\hbar \langle n|[\mathcal{H}_{\bm E},\rho]\ket{p} = e\bm{E}(t).\big[\mathcal{D}_{\bm{k}}\rho\big]^{np}$, where $\big[\mathcal{D}_{\bm{k}}\rho\big]^{np} = \partial_{\bm {k}} \rho^{np}-i[\mathcal{R}_{\bm{k}},\rho]^{np}$ is a covariant derivative~\cite{Nagaosa_AM2017, Bhalla_prb2023}, where $\partial_{\bm k}\equiv \partial/\partial \bm k$ .  The Berry connection in the $\bm{k}$ space is denoted by $\mathcal{R}_{\bm{k}}^{np}=\bra{u_{\bm{k}}^n}\ket{i\partial_{\bm{k}}u_{\bm{k}}^p}$.
The third term $\mathcal{J}_{\bm k}[\rho]=i/\hbar[U,\rho]$, takes care of the effect of disorder potential $U$ in the system. In this study, we consider weak disorder and treat it within the first order Born approximation. The latter is defined in the band basis representation like
\begin{align}\label{eqn:J}\nonumber
\mathcal{J}_{\bm{k}}^{np}[\rho]=\frac{1}{\hbar^2}\sum_{\bm{k}'}\sum_{qt}\bigg[\frac{U_{\bm{kk}'}^{nq}U_{\bm{k}'\bm{k}}^{qt}\space\rho^{qp}_{\bm{k}}}{i\big(\varepsilon_{\bm{k}'}^q-\varepsilon_{\bm{k}}^t\big)/\hbar}-\frac{U_{\bm{kk}'}^{nq}U_{\bm{k}'\bm{k}}^{tp}\space\rho^{qt}_{\bm{k}'}}{i\big(\varepsilon_{\bm{k}'}^t-\varepsilon_{\bm{k}}^p\big)/\hbar}\\
-\frac{U_{\bm{kk}'}^{nq}U_{\bm{k}'\bm{k}}^{tp}\space\rho^{qt}_{\bm{k}'}}{i\big(\varepsilon_{\bm{k}'}^n-\varepsilon_{\bm{k}}^q\big)/\hbar}+\frac{U_{\bm{kk}'}^{qt}U_{\bm{k}'\bm{k}}^{tp}\space\rho^{nq}_{\bm{k}}}{i\big(\varepsilon_{\bm{k}'}^q-\varepsilon_{\bm{k}}^t\big)/\hbar}\bigg].
\end{align}
In the band basis representation, the averages of the product of two elements of the disordered matrix are expressed in the form $U_{\bm{kk}'}^{nq} U_{\bm{k}'\bm{k}}^{qt} = U_0^2 \langle u_{\bm{k}}^n | u_{\bm{k}'}^q \rangle \langle u_{\bm{k}'}^q | u_{\bm{k}}^t \rangle$. Here, $U_{\bm{kk}'}^{nq} = \langle n, \bm{k} | U | q, \bm{k}' \rangle$ is the disorder potential matrix. Here, the average of disorder potential is zero $<U_{\bm k \bm k'}^{np}>=0$. The second-order spatial correlation only depends on the difference in positions~\cite{Yang_prb2011}. 
\begin{figure*}[htp]
    \centering
    \includegraphics[width=15cm]{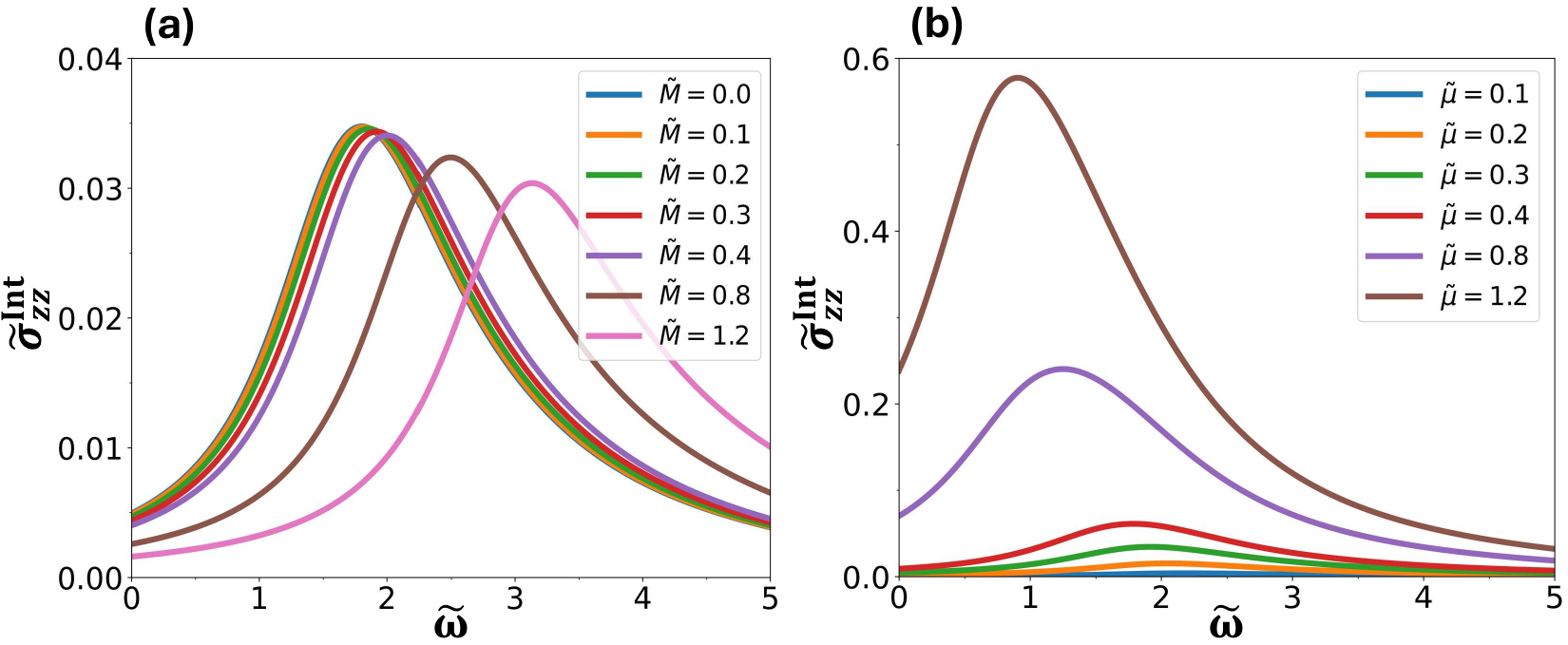}
    \caption{(a) and (b) show the intrinsic contribution to the interband part of longitudinal response of DNLSM ($\tilde{\sigma}_{zz}^{\text{Int}}=\sigma_{zz}^{\text{Int}}/\sigma_0$) with varying frequency at different mass values (at constant $\tilde{\mu}=0.3$) and the different chemical potential values (at constant $\tilde{M}=0.3$).} 
    \label{fig:Intrinsic}
\end{figure*}

Further,  we express the density matrix in the band basis representation by taking the field correction terms as: $\rho_{\bm k}=\rho^{nn}_{0, \bm k}+\mathcal{N}^{nn}_{\bm E, \bm k} + S^{np}_{\bm E, \bm k}$. The term $\rho^{nn}_{0, \bm k}$ is the equilibrium Fermi Dirac distribution function, defined as $\rho^{nn}_{0, \bm k} = f^{0}(\varepsilon_{\bm k}^{n})=[1+e^{\beta(\varepsilon_{\bm k}^{n}-\mu)}]^{-1}$, where $\beta=[k_{B}T]^{-1}$ is the energy where $k_{B}$ is the Boltzmann constant, $T$ the absolute temperature associated with the electron and $\mu$ the chemical potential. The electric field dependent terms of the density matrix $\mathcal{N}^{nn}_{\bm E, \bm k}$ and $S^{np}_{\bm E, \bm k}$ refer to intraband (diagonal) and interband (off diagonal) part of the density matrix respectively. Following the quantum kinetic equation for intraband part, we have $\mathcal{N}_{\bm E,{\bm{k}}}^{nn}=\frac{e\space\bm{E}}{ (g_{\bm k}+i\hbar\omega)} .\frac{\partial\rho^{nn}_{0,\bm{k}}}{\partial \bm{k}}$, where $g_{k} = \hbar/\tau_{\bm k}$, $\tau_{\bm k}$ is the relaxation time associated with the intraband scattering events. On the other hand, the interband part $S^{np}_{\bm E, \bm k}$ gives $ S_{\bm E,{\bm{k}}}^{np}= \frac{\hbar{D_{\bm E,{\bm{k}}}^{np}} + \hbar \mathcal{J}_{\bm{k}}^{np}[\mathcal{N}_{\bm E,{\bm{k}}}]}{g+\space i\space  (\omega^{np}+\hbar  \omega)}$
, $g = \hbar/\tau$ with $\tau$ as the interband relaxation time and $\hbar\omega^ {pn}= \varepsilon_{{\bm k}}^{p}-\varepsilon_{{\bm k}}^{n}$ is the energy difference between the $p$ and $n$ bands~\cite{Pandey_prb2024}. From here, we infer that ${D_{\bm E,{\bm{k}}}^{np}}$ leads to the field driven (intrinsic) part and $\mathcal{J}_{\bm{k}}^{np}[\mathcal{N}_{\bm E,{\bm{k}}}]$ to the scattering driven (extrinsic) part to the interband contribution of the total conductivity.

\section{Dynamical interband conductivity in DNLSMs}

On application of the field, the generation of an electrical current can be mathematically deduced by using relation $\bm J=\text{Tr}[\bm v \rho]$, where $\bm v$ stands for the velocity of an electron and $\sigma$ is the conductivity tensor.
In the band basis representation, the normalised band velocity can be expressed in the form $ \space \tilde{v}^{pn}_i=\space\delta_{pn}\space\partial_{\tilde{k}_i}\tilde{\varepsilon}_{\bm{{k}}}^n+i\space\mathcal{\tilde{R}}^{pn}_{{\bm k}_i}\space\tilde{\omega}^{pn}$, where $\tilde{v}_i=\hbar  {k}_0 v_i/\varepsilon_0$ and $\tilde{\omega}^{pn}=\hbar \omega^{np}/\varepsilon_0$. As discussed earlier, the $\rho$ comprises the intraband and interband contributions via correction terms $\mathcal{N}^{nn}_{\bm E, \bm k}$ and $S^{np}_{\bm E, \bm k}$ respectively. These terms further lead to the conductivity, reckoned on following the relation  $\text{Tr}[\bm v \rho]=\sigma \bm E$. 
Depending on the direction of field, one can expect various components of the conductivity tensor. However, all the anomalous terms for DNLSMs vanish such as $\sigma_{xy} = \sigma_{yx} = \sigma_{zx}= \sigma_{xz}= \sigma_{zy} =\sigma_{yz} = 0$, 
this happens due to the presence of the inversion ($\mathcal{P}$)-symmetry points (in the nodal ring), net Hall current vanishes because it contributes to the transverse current with opposite signs~\cite{Wang_prb2021}. Hence, one requires an application of $\mathcal{P}$-symmetry broken term to generate a non-zero Hall current. 
Therefore, we are left out with the longitudinal components such as $\sigma_{xx},\sigma_{yy} $ and $\sigma_{zz}$. Interestingly, the only disorder contribution, termed as extrinsic originating from $\mathcal{J}_{\bm{k}}^{np}[\mathcal{N}_{\bm E,{\bm{k}}}]$, associated with $\sigma_{zz}$ remains finite and other $\sigma_{xx}$ and $\sigma_{yy}$ go to zero due to the nature of the band velocity in the respective directions. This motivates the investigation for the longitudinal response $\sigma_{zz}$ due to the field along $z$-direction and the disorder contribution to the net conductivity. Further, to capture the feature beyond the traditional Drude like behavior of intraband response, the study of interband part becomes more relevant and significant.  
Additionally, the symmetric density of states for the NLSMs due to preserved symmetries makes the study of optical conductivity more interesting in comparison to Weyl semimetals having asymmetric density of states due to breaking of either inversion or time reversal symmetry.

The total interband conductivity is the addition of the extrinsic and intrinsic contributions of the conductivity.
First, the extrinsic contribution to the interband part of longitudinal conductivity gives 
\begin{align}
\label{eqn:ext_zz}
\sigma_{zz}^{\text{Ext}} = - \frac{ie}{ E_z} \sum_{n\neq p}\sum_{\tilde{\bm k}}\tilde{\mathcal{R}}_{{ k}_z}^{np}\frac{\mathcal{J}_{\tilde{{k}}_z}^{np}[\mathcal{N}_{\bm E, \bm k}]\tilde{\omega}^{pn}}{\tilde{g}+\space i\space (\tilde{\omega}^{pn}+ \tilde{\omega})}.
\end{align}
Here $\tilde{g}=\frac{\hbar}{\varepsilon_0 \tau}$, $\tilde{\omega}=\hbar \omega/\varepsilon_0$ and $\tilde{F}^{np}=f^0{(\tilde{\varepsilon}_{\bm{k}}^n)}-f^0{(\tilde{\varepsilon}_{\bm{k}}^p)}$ refer to the occupation probability difference between two separate bands. 
 From~\eqref{eqn:ext_zz}, we find $\sigma_{zz}^{\text{Ext}} \propto \mathcal{\tilde{R}}_{k_z}^{np} \mathcal{J}_{\tilde{{k}}_z}^{np}[\mathcal{N}_{\bm E, \bm k}] $ at $\tilde{g}=0$ and $\tilde{\omega}=0$. Further to explore the variation of extrinsic part of conductivity in the different regimes of chemical potential ($\tilde{\mu}$), mass ($\tilde{M}$) and frequency ($\tilde{\omega}$), we present a plot for the extrinsic interband response in figure~\ref{fig:Extrinsic}.

 figure~\ref{fig:Extrinsic}(a) depicts the variation of the extrinsic interband part of the conductivity with the frequency $\tilde{\omega}(\omega/\varepsilon_0)$ at distinct values of the mass term $\tilde{M}(M/\varepsilon_0)$. Here we fixed the chemical potential as $\tilde{\mu}=0.3$.
We find the extrinsic contribution generates a peak as the frequency approaches twice the chemical potential i.e., $\tilde{\omega}=2 \tilde{\mu}$. Furthermore, the location of the peak remains intact at different values of mass.
Additionally, the strength and sign of the extrinsic response rely on the limiting regimes of $\tilde{M}$ and $\tilde{\mu}$. The whole picture occurs due to the shift of the electron's distribution function in the momentum space. This assures the dominant variation at the band energy equals to the chemical potential within the low temperature range. Also, when the gap value matches the chemical potential, a sign transition in the extrinsic response happens.  

\begin{figure*}[htp]
    \centering
    \includegraphics[width=15cm]{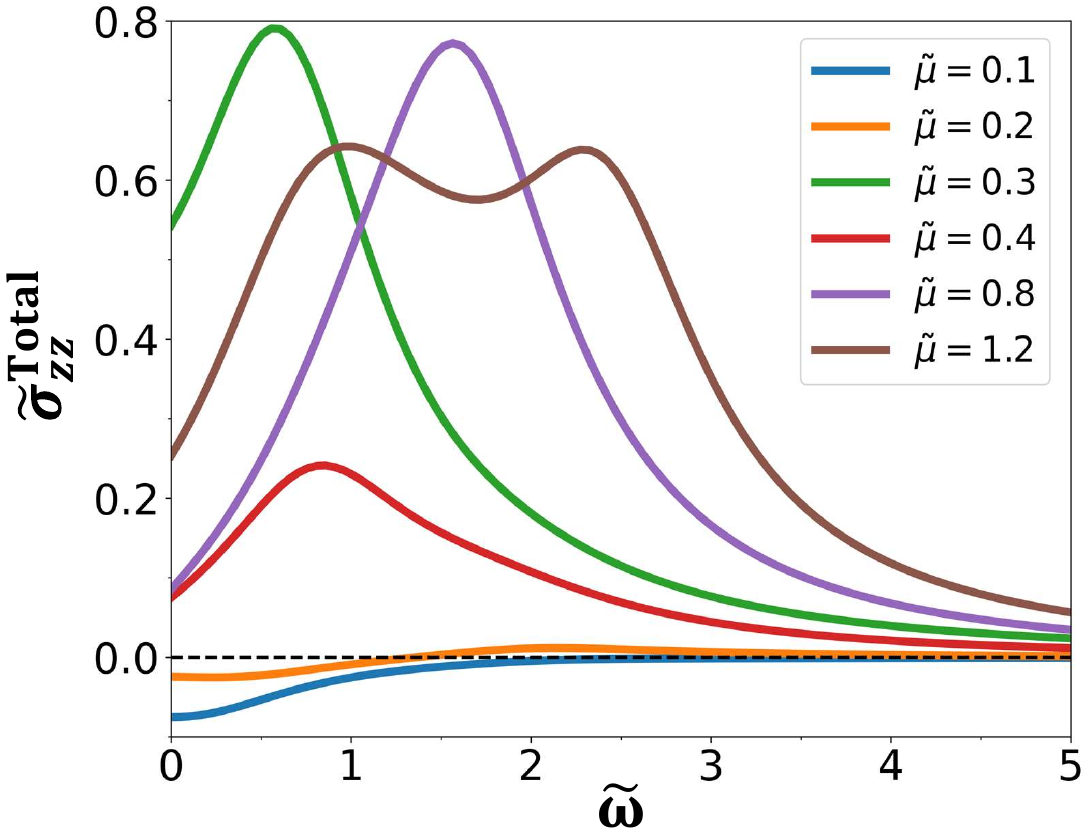}
    \caption{Plot for the variation of total interband part of longitudinal conductivity of DNLSM ($\tilde{\sigma}_{zz}^{\text{Total}}=(\sigma_{zz}^{\text{Int}}+\sigma_{zz}^{\text{Ext}})/\sigma_0$) versus frequency at different chemical potential values and keeping fixed $\tilde{M}=0.3$.} 
    \label{fig:Total}
\end{figure*}

For better understanding, we show $\tilde{\sigma}_{zz}^{\text{Ext}}$ with $\tilde{\omega}$ at different values of chemical potential $\tilde{\mu}$ and fixed $\tilde{M}=0.3$ in figure~\ref{fig:Extrinsic}(b).
Clearly, we can see the location of the peak is decided by the value of the chemical potential as stated earlier. Further, the extrinsic response gives a finite value at zero frequency and approaches towards zero value at high frequency.
Second, the intrinsic interband part of the longitudinal conductivity is
\begin{align}\label{eqn:int_zz}
\sigma_{zz}^{\text{Int}} =\frac{e^2}{\hbar}  \sum_{n\neq p}\sum_{ \tilde{\bm k}}  \frac{|\mathcal{\tilde{R}}^{pn}_{{k}_{z}}|^2 \tilde{F}^{np} \tilde{\omega}^{pn}}{\tilde{g}+\space i \space ( \tilde{\omega}^{pn} + \tilde{\omega})}.
\end{align}
From this expression, we find the response decays to zero value at high frequency. On the contrary, at zero frequency it gives a finite value with smaller magnitude. Interestingly, the intrinsic interband response figure~\ref{fig:Intrinsic}, shows exotic behavior at intermediate frequency values for different mass and chemical potential. Here, the location of the peaks moves towards a higher frequency value with increasing $\tilde{M}$ and toward a lower frequency value with increasing $\tilde{\mu}$. 
Here, the occurrence of the peak originates via the factor $\tilde{F}^{np}$ in equation~\eqref{eqn:int_zz}, transforms into the Heaviside function at the low temperature like $\theta(\tilde{\varepsilon}_{\bm k}-\tilde{\mu})$ or $\theta(\mathcal{\tilde{K}}-\mathcal{\tilde{K}}_F)$. Depending on the chemical potential taken into account, which gives the inputs for integration limits, consideration of the number of states to compute response changes and hence the location of the peak moves.  
\begin{table*}[htp]
    \centering
    \fontsize{7pt}{7pt}\selectfont  
    \renewcommand{\arraystretch}{1.5}  
    \setlength{\arrayrulewidth}{0.5pt} 
    \begin{tabular}{|>{\centering\arraybackslash}p{0.7cm}|>{\centering\arraybackslash}p{2.0cm}|>{\centering\arraybackslash}p{1.5cm}|>{\centering\arraybackslash}p{1.5cm}|>{\centering\arraybackslash}p{2.5cm}|>{\centering\arraybackslash}p{2.5cm}|>{\centering\arraybackslash}p{2.5cm}|} 
    \hline  
    \textbf{S.No.} & $\hbar\omega(meV)$& $\mu(meV)$& $M(meV)$& \textbf{$\sigma^{\text{Int}}_{zz}$}$(e^2/h)$& \textbf{$\sigma^{\text{Ext}}_{zz}$}$(e^2/h)$& \textbf{$\sigma^{\text{Total}}_{zz}$}$(e^2/h)$\\
    \hline  
    1 & 90& 50& 50& 0.007453& 0.955464& 0.962917\\
    \hline  
    2 & 90& 50& 150& 0.003798& -0.016407& -0.012609\\
    \hline   
    3 & 90& 150& 50& 0.195005& 0.168391& 0.363396\\
    \hline    
    4& 180& 50& 50& 0.013959& 0.661144& 0.675103\\
    \hline  
    5& 180& 50& 150& 0.006102& -0.011046& -0.004944\\
    \hline  
    6& 180& 150& 50& 0.302351& 0.430628& 0.732979\\\hline  
    \end{tabular}
    \caption{Table showing numerical estimation of the total longitudinal AC interband response ($\sigma_{zz}^{\text{Total}}$)  of $\text{Ca}_3\text{P}_2$ on the application of an oscillating electric field in the $z$ direction, which is a combination of the intrinsic ($\sigma_{zz}^{\text{Int}}$)  and extrinsic ($\sigma_{zz}^{\text{Ext}}$) components at different values of the parameters such as $\hbar\omega$, $\mu$ and $M$.}
    \label{tab:SP}
\end{table*}

In figure~\ref{fig:Total}, we show the total interband response $\tilde{\sigma}^{\text{Total}}_{zz}$, addition of the intrinsic and extrinsic contributions, for DNLSM with frequency $\tilde{\omega}$ at distinct values of the chemical potential. We observe the dominance of the extrinsic interband part over the intrinsic part in the low-frequency to mid-frequency regimes. However, at $\tilde{\mu}=0.8$ the competition between the extrinsic and intrinsic begins, which further generates two separate peaks associated with both parts at $\tilde{\mu}=1.2$. 
For comparative analysis, the numerical estimation of both extrinsic and intrinsic contributions to the interband part of longitudinal response for DNLSM Ca$_3$P$_2$ are provided in Table~\ref{tab:SP} at distinct parameter values. Based on the density functional theory (DFT), the parameters for $\text{Ca}_3\text{P}_2$ are considered as the radius of nodal ring $k_0 \approx 0.206 \AA^{-1}$, energy associated with ring radius $\varepsilon_0 \approx 0.184 eV$ and $\gamma \approx 2.80$ ~\cite{Chan_prb2016}. Moreover, the magnitude of the interband $\tilde{\sigma}_{zz}$ shown in~\cite{Barati_prb2017} is comparable to the magnitude of $\tilde{\sigma}_{zz}^{\text{Int}}$ observed in our calculation. 

On similar ground the interband response of the two-dimensional nodal line semimetal can also be examined. Here, the interband response will solely arise from the intrinsic (field-driven) part of the conductivity, while the extrinsic (scattering driven) contribution vanishes due to the two dimensional nature of the system. Further, the location of the characteristic peak and the strength of the interband conductivity depending on the competition between the energy of the system and the applied frequency will be affected. 

\section{Experimental relevance and validity}
\label{Sec:Experimental relevance and validity}

Experimentally, the longitudinal conductivity for the DNLSM can be investigated by measuring the current. The three dimensional sample is placed and connected to a voltage source,  with a bias applied along the $z$ direction, and the current is measured in the same direction. To modulate and detect intrinsic and extrinsic contributions, the chemical potential can be tuned via gating techniques or electronic doping. On applying bias $x$ and $y$ directions, one can get only contributions related to the field driven part.

The present work is valid in the limit $\mu \gg \hbar/\tau$ (i.e., the condition for the weak disorder case). In this limit, a small modification via disorder potential is introduced in the total Hamiltonian of the system. To treat the related effects, we have used the first-order Born approximation and retained finite-order terms according to the disorder's strength. The considered approximations are influential when disorder levels are weak to moderate; however, as disorder increases, the validity of the first-order Born approximation diminishes.
For strong disorder limits $\mu \ll \hbar/\tau$, the higher-order corrections within the scattering term become prominent, requiring methods beyond the first-order Born approximation due to the failure of the Bloch band structure or by renormalizing the density matrix via standard spectral function approach.
Additionally, in the presence of weak disorder, the nodal line is well defined in the momentum space (slightly broadened), and the system remains gapless as long as the symmetry is protected. 
In contrast, in the case of strong disorder, the band broadening happens and the topological features become ill-defined (not sharply defined).
Furthermore, our calculations are limited for the low temperature regime. At high temperatures, the other scattering events such as electron-phonon scattering may play a role which may modify the overall response of the system.

\section{Summary}
\label{sec: summary}
In summary, we have investigated the dynamical longitudinal interband part of conductivity for three dimensional DNLSMs, which is a $\mathcal{PT}$ symmetry-broken system, under the influence of the time dependent electric field and the presence of the weak disorder by using the quantum kinetic approach.
The total longitudinal interband response of a system has two components, field driven part or intrinsic and scattering driven part or extrinsic.
The analysis has been performed following the DFT fitted model Hamiltonian for $\text{Ca}_3\text{P}_2$ three dimensional system.

The extrinsic interband part shows the location of observed peak depending on both the chemical potential and the mass value. The increase in the chemical potential causes the peak to shift towards the lower frequency regime and the increase in mass value shifts the peak towards the higher frequency regime. On the other hand, the resonant peak corresponds to the intrinsic interband contribution, which depends only on the chemical potential. We also found that the scattering driven response dominates over the field driven part in the low frequency regime, thus enhancing the overall response of the system. The presented work can be extended for other nodal line systems by taking into account tilt, strain, etc., which will be beneficial for future device application. In addition, one can also extend the present study by taking into account the electron-phonon interaction.
\section*{Acknowledgment}
This work is financially supported by the Science and Engineering Research Board-State University Research Excellence under project number SUR/2022/000289. 
\twocolumngrid
\bibliography{Ref}

\end{document}